\documentstyle[aps,prd]{revtex} 
\begin{document} 
%%%%%%%%%%%%%%%%%%%%%%%%%%%%%%%%%%%%%%%%%%%%%%%%%%%%%%%%%%%%%%%%%%%%%% 
\title{Gravitational vacuum polarization II:\\
Energy conditions in the Boulware vacuum\\
gr-qc/9604008} 
\author{Matt Visser\cite{e-mail}} 
\address{Physics Department, Washington University, St. Louis,  
         Missouri 63130-4899} 
\date{1 April 1996} 
\twocolumn[
\maketitle
%%%%%%%%%%%%%%%%%%%%%%%%%%%%%%%%%%%%%%%%%%%%%%%%%%%%%%%%%%%%%%%%%%%%%%
\parshape=1 0.75in 5.5in
%%%\maketitle 
%%%\begin{abstract} 
Building on techniques developed in an earlier paper, I investigate
the various point-wise and averaged energy conditions for the
quantum stress-energy tensor corresponding to a conformally coupled
massless scalar field in the Boulware vacuum. I work in the test-field
limit, restrict attention to the Schwarzschild geometry, and invoke
a mixture of analytical and numerical techniques.  In contradistinction
to the case of the Hartle--Hawking vacuum, wherein violations of
the energy conditions were confined to the region between the event
horizon and the unstable photon orbit, {\bf I show that in the
Boulware vacuum (1) all standard (point-wise and averaged) energy
conditions are violated throughout the exterior region---all the
way from spatial infinity down to the event horizon, and (2)
outside the event horizon the standard point-wise energy conditions
are violated in a maximal manner: they are violated at all points
and for all null/timelike vectors.}  (The region inside the event
horizon is considerably messier, and of dubious physical relevance.
{\bf Nevertheless the standard point-wise energy conditions seem
to be violated even inside the event horizon.})  I argue that this
is highly suggestive evidence pointing to the fact that general
self-consistent solutions of semiclassical quantum gravity might
{\em not} satisfy the energy conditions, and may in fact for certain
quantum fields and certain quantum states violate {\em all} the
energy conditions.
%%%\end{abstract}
\vskip 0.125 in
\parshape=1 0.75in 5.5in
PACS number(s): 04.20.-q    04.20.Gz    04.25.-g    04.90.+e
\pacs{}
]
 
%%%%%%%%%%%%%%%%%%%%%%%%%%%%%%%%%%%%%%%%%%%%%%%%%%%%%%%%%%%%%%%%%%%%%%%%%% 
\pacs{04.60.+v 04.70.Dy} 
%%%%%%%%%%%%%%%%%%%%%%%%%%%%%%%%%%%%%%%%%%%%%%%%%%%%%%%%%%%%%%%%%%%%%%%%%% 
%%%\bibliographystyle{prsty} 
%%%\bibliographystyle{unsrt} 
%%%%%%%%%%%%%%%%%%%%%%%%%%%%%%%%%%%%%%%%%%%%%%%%%%%%%%%%%%%%%%%%%%%%%%%%%% 
%%%\newtheorem{theorem}{Theorem} 
%%%\newtheorem{definition}{Definition} 
%%%%%%%%%%%%%%%%%%%%%%%%%%%%%%%%%%%%%%%%%%%%%%%%%%%%%%%%%%%%%%%%%%%%%%%%%% 
\narrowtext 
\section{INTRODUCTION} 
%%%%%%%%%%%%%%%%%%%%%%%%%%%%%%%%%%%%%%%%%%%%%%%%%%%%%%%%%%%%%%%%%%%%%%%%%% 
 
Investigations of the gravitationally induced vacuum polarization
produced when a quantum field theory is constructed on a curved
background spacetime are a topic of considerable current
interest~\cite{Visser96a,Visser95,Visser,Flanagan-Wald,%
Ford-Roman94,Ford-Roman93,Ford-Roman96,Ford-Roman95b}. A key aspect
of these investigations is the manner in which the various energy
conditions are affected by this gravitational vacuum polarization.
This general topic is of critical importance to attempts to generalise
the classical singularity theorems~\cite{Hawking-Ellis}, classical
positive mass theorems~\cite{Penrose-Sorkin-Woolgar}, and classical
laws of black hole dynamics to semiclassical quantum
gravity~\cite{Visser95,Visser}.

It is perhaps a little sobering to realise that none of the currently
known versions of these classical theorems survive the introduction
of even semiclassical quantum gravity, let alone full-fledged
quantum gravity~\cite{Visser95,Visser}.

In this paper I shall use techniques developed in an earlier
paper~\cite{Visser96a} to explore these issues in a little more
detail: I restrict attention to the conformally coupled massless
scalar field on a Schwarzschild spacetime, in the Boulware vacuum.
I shall continue to work in the test-field limit.

For this geometry and vacuum state one has both (1) a useful analytic
approximation to the gravitational polarization---obtained by
combining the Page approximation for the Hartle--Hawking
vacuum~\cite{Page82} with the Brown--Ottewill
approximation~\cite{Brown-Ottewill} for the difference between the
Hartle--Hawking and Boulware vacua, and (2) numerical estimates of
the vacuum polarization---these estimates being obtained by combining
the numerical calculations of Howard~\cite{Howard}, and Howard and
Candelas~\cite{Howard-Candelas},  (who calculate the stress-energy
tensor in the Hartle--Hawking vacuum), with the further numerical
calculations of Jensen, McLaughlin, and Ottewill~\cite{JLO92} (who
numerically calculate the difference between the Hartle--Hawking
and Boulware vacua). Further refinements using the numerical data
of and Anderson, Hiscock, and Samuel~\cite{AHS,AHS-2,Anderson-Private}
are certainly possible, but this avenue has not yet been explored.

For the Hartle--Hawking vacuum I found that the various energy
conditions were violated in a  nested set of onion-like layers
located between the event horizon and the unstable photon
orbit~\cite{Visser96a}. Furthermore many of the energy conditions
continued to be violated inside the event horizon.

For the Boulware vacuum the situation is even easier to describe:
\begin{itemize}
\item
{\bf (1) All the standard point-wise energy conditions and standard
averaged energy conditions conditions are violated throughout the
entire region exterior to the event horizon---all the way from
spatial infinity down to the event horizon.}
\item
{\bf (2) Outside the horizon, the standard point-wise energy
conditions are violated in a maximal manner:  they are violated at
all points and for all null/timelike vectors.}
\item
{\bf (3) The standard point-wise energy conditions seem to be
violated even inside the event horizon.}
\end{itemize}
This includes the obvious (point-wise) null,
weak, strong, and dominant energy conditions (NEC, WEC, SEC, and
DEC), the averaged null, weak, and strong energy conditions (ANEC,
AWEC, and ASEC), more exotic energy conditions such as the partial
null energy  condition (PNEC), the asymptotic null energy condition
(Scri--NEC), the averaged asymptotic null energy condition
(Scri--ANEC), as well as various one-sided integral averages and
averages constructed by allowing arbitrary weighting along the
curve of interest. (For definitions of the basic energy conditions
see~\cite{Hawking-Ellis}, or~\cite{Visser95}.  For definitions of
some of the more exotic energy conditions see~\cite{Visser96a}.)

Energy conditions {\em not} completely pinned down by this type of
analysis include the ``quantum inequalities'' of Ford and
Roman~\cite{Ford-Roman93,Ford-Roman96,Ford-Roman95b}, and versions
of the ANEC in which one is willing to countenance some form of
bounded negativity for the ANEC integral.

The situation inside the event horizon is rather messier. Calculations
based on the analytic approximation indicate that some types of
energy condition become satisfied sufficiently close to the
singularity. On the other hand it should be borne in mind that the
stress-tensor for the Boulware vacuum is singular at the event
horizon---thus there are good reasons for not taking the Boulware
vacuum state too seriously inside the event horizon.

The key results underlying this wholesale violation of the
energy-conditions are:\\
%%%\begin{theorem}[NEC/WEC/SEC/DEC violation]\hfil\break
{\bf Theorem  1 [NEC/WEC/SEC/DEC violation]}\hfil\break
\label{t-local}
{\em 
For a conformally-coupled massless scalar field on a Schwarzschild
spacetime in the Boulware vacuum state:  For {\em any} point $p$
anywhere outside the event horizon {\em there exist} null and
timelike vectors such that
\begin{eqnarray}
&\langle B | T^{\mu\nu} | B \rangle& \; k_\mu k_\nu \; \leq 0.
\nonumber\\
&\langle B | T^{\mu\nu} | B \rangle& \; V_\mu V_\nu \; \leq 0.
\nonumber\\
&\langle B | \bar T^{\mu\nu} | B \rangle& \; V_\mu V_\nu \; \leq 0.
\end{eqnarray}
(Here $\bar T$ is the trace-reversed stress-energy tensor.) 
Thus NEC, WEC, SEC, and DEC are all violated outside the event horizon.
}
%%%\end{theorem}

If one is willing (with suitable caveats to be described below) to
accept the analytic approximation to the stress-energy tensor as
a reliable guide then the above result may be extended to the entire
spacetime.

Outside the horizon, one can in fact prove a much stronger result:\\
%%%\begin{theorem}[Total external NEC violation]\hfil\break
{\bf Theorem 2 [Total external NEC violation]}\hfil\break
\label{t-total}
{\em
For a conformally-coupled massless scalar field on a Schwarzschild
spacetime in the Boulware vacuum state:  For {\em any} point $p$
outside the event horizon and {\em any} null vector $k$
\begin{equation}
\langle B | T^{\mu\nu} | B \rangle \; k_\mu k_\nu \; \leq 0.
\end{equation}
The equality is in fact achieved only at spatial infinity.
}\\
%%%\end{theorem}
Similar results can be proved for the other standard point-wise
energy conditions. 

Even if one is willing (with suitable caveats) to accept the analytic
approximation to the stress-energy tensor as a reliable guide then
this second theorem does {\em not} extend to the entire spacetime.

%%%%%%%%%%%%%%%%%%%%%%%%%%%%%%%%%%%%%%%%%%%%%%%%%%%%%%%%%%%%%%%%%%%%%%%%%%%%%% 
\section{Vacuum polarization in Schwarzschild spacetime: 
         Boulware vacuum} 
%%%%%%%%%%%%%%%%%%%%%%%%%%%%%%%%%%%%%%%%%%%%%%%%%%%%%%%%%%%%%%%%%%%%%%%%%%%%%% 
 
By spherical symmetry one knows that 
\begin{equation} 
\langle B | T^{\hat\mu}{}_{\hat\nu} | B \rangle \equiv  
\left[ \matrix{-\rho&0&0&0\cr 
               0&-\tau&0&0\cr 
               0&0&+p&0\cr 
	       0&0&0&+p\cr } \right]. 
\end{equation} 
Where $\rho$, $\tau$ and $p$ are functions of $r$, $M$ and $\hbar$.
(Note: I set $G\equiv1$, and choose to work in a local-Lorentz
basis attached to the fiducial static observers [FIDOS].)

A subtlety arises when working in a local-Lorentz basis and looking
at the two-index-down (or two-index-up) versions of the stress-energy.
Outside the horizon one has
\begin{equation} 
g^{\hat\mu\hat\nu}|_{\rm outside} =  
\left[ \matrix{-1&0&0&0\cr 
               0&+1&0&0\cr 
               0&0&+1&0\cr 
	       0&0&0&+1\cr } \right]. 
\end{equation} 
Consequently
\begin{equation} 
\langle B | T^{\hat\mu\hat\nu} | B \rangle |_{\rm outside} = 
\left[ \matrix{+\rho&0&0&0\cr 
               0&-\tau&0&0\cr 
               0&0&+p&0\cr 
	       0&0&0&+p\cr } \right]. 
\end{equation} 
Inside the horizon on the other hand, it is the radial direction
that is timelike, so
\begin{equation}
g^{\hat\mu\hat\nu} |_{\rm inside}=  
\left[ \matrix{+1&0&0&0\cr 
               0&-1&0&0\cr 
               0&0&+1&0\cr 
	       0&0&0&+1\cr } \right]. 
\end{equation} 
Consequently one has the potentially confusing result that
\begin{equation} 
\langle B | T^{\hat\mu\hat\nu} | B \rangle|_{\rm inside}  =
\left[ \matrix{-\rho&0&0&0\cr 
               0&+\tau&0&0\cr 
               0&0&+p&0\cr 
	       0&0&0&+p\cr } \right]. 
\end{equation} 
Thus, inside the horizon, one should interpret $\tau$ as the energy
density and $\rho$ as the tension (this tension now acting in the
spacelike $t$-direction).

Recall that if one wishes the density as measured by a freely-falling
observer to remain finite as one crosses the event horizon then
one needs $\rho=\tau$ at $r=2M$. This condition is most definitely
not satisfied by the Boulware vacuum and leads to discontinuous
behaviour of the energy conditions as one crosses the horizon. This
is not particularly surprising since the Boulware stress-energy is
itself singular at the event horizon. I shall put aside for now
the worry that the Boulware vacuum might be completely nonsensical
inside the event horizon and do as much as is currently possible
by using the analytic approximation in that region.

To start the actual analysis I require explicit analytic (though
approximate) formulae for the stress-energy tensor. By combining
Page's analytic approximation~\cite{Page82} for the Hartle--Hawking
vacuum state with the Brown--Ottewill~\cite{Brown-Ottewill} analytic
approximation for the difference between the Boulware and
Hartle--Hawking vacua one obtains a simple rational polynomial
approximation to the stress--energy tensor:
\begin{eqnarray} 
\rho(r) &=&  
-3 p_\infty \; ({2M/r})^6 \;
{ [40 - 72 (2M/r) + 33 (2M/r)^2] \over (1 - 2M/r)^2 }, 
\\ 
\tau(r) &=& 
- p_\infty  \;  ({2M/r})^6 \;
{ [8 - 24 (2M/r) + 15 (2M/r)^2] \over (1 - 2M/r)^2 },
\\ 
p(r)    &=& 
- p_\infty \;  ({2M/r})^6  \;
{ [4-  3(2M/r)]^2 \over (1 - 2M/r)^2 }. 
\end{eqnarray} 
Here I have defined a constant 
\begin{equation} 
p_\infty \equiv {\hbar\over 90 (16\pi)^2 (2M)^4}.  
\end{equation} 
In the Hartle--Hawking vacuum $p_\infty$ can be interpreted as
the pressure at spatial infinity.

To get these expressions I have explicitly expanded the functions
given in references~\cite{Page82,Howard} as polynomials in $2M/r$
to obtain the formulae for the Hartle--Hawking vacuum given
in~\cite{Visser96a}, the results being checked against
Elster~\cite{Elster83} and the spin-zero case of
Brown--Ottewill~\cite{Brown-Ottewill}. (Combine equations (3.11)
and (3.12) on page 2517.)

Next the spin-zero Brown--Ottewill analytic approximation for the
difference term is evaluated~\cite{Brown-Ottewill}
\begin{eqnarray}
&&
\langle B | T^{\hat\mu}{}_{\hat\nu} | B \rangle  - 
\langle H | T^{\hat\mu}{}_{\hat\nu} | H \rangle 
= 
\nonumber\\
&&\qquad\qquad
- p_\infty \; {1\over(1-2M/r)^2} \;
\left[ \matrix{-3&0&0&0\cr 
               0&1&0&0\cr 
               0&0&1&0\cr 
	       0&0&0&1\cr } \right]. 
\end{eqnarray}
(Combine equations (3.16) and (3.17) on page 2517. Set $T=0$
for the Boulware vacuum and note that the coefficients of their
$V^{\mu\nu}$ term cancel for spin zero.)

The history of this last expression is interesting: This expression was
first written down by Christensen and Fulling who conjectured that
this result was {\em exact}~\cite[equation (6.29) page
2101]{Fulling77}.  Later on, Brown and Ottewill effectively
derived this result as an {\em approximation} in their analytic
approximation scheme. Finally, Jensen, McLaughlin, and Ottewill
showed that this result is not in fact exact, but is nevertheless
a good approximation~\cite{JLO92}.

As a consistency check, the trace of the stress--energy tensor is
given by
\begin{equation} 
\langle B | T^{\hat\mu}{}_{\hat\mu} | B \rangle \equiv  
-\rho - \tau +2 p \equiv 
96 \; p_\infty \; (2M/r)^6. 	  
\end{equation} 
This result, because it is simply a restatement of the conformal 
anomaly, is known to be exact. 

I intend to use this analytic approximation, subject to suitable
caveats, over the entire maximally extended Kruskal--Szekeres
manifold. (That is, over the entire spacetime of an eternal black
hole).
 
Outside the event horizon, the explicit numerical computations of
Jensen, McLaughlin, and Ottewill~\cite{JLO92} show that this analytic
approximation is in good qualitative agreement with the numerically
determined stress-energy tensor.

Several general results can immediately be extracted from this
analytic approximation:
\begin{itemize}
\item
The quantity $\rho$ is negative over the entire range $r\in[0,\infty]$.
Outside the event horizon you should interpret this as the energy
density, inside the event horizon it is the tension.
\item
%%%(
The quantity $\tau$ is negative in the range $r\in
[0,1.7753M)\cup(4.2274M,\infty]$, and positive in the range
$r\in(1.7753M,4.2274M)$. Outside the event horizon you should
interpret this as the radial tension, inside the event horizon it
is the energy density.
%%%)
\item
The transverse pressure $p$ is negative over almost the entire
range $r\in[0,\infty]$, with the exception of $r=3M/2$ where it is
zero.
\end{itemize}

The numerical data~\cite{JLO92} is limited to the region outside
the event horizon $r\in[2M,\infty]$. By visual inspection of the
graphs one draws the general conclusions that:
\begin{itemize}
\item
The density $\rho$ is negative over the entire range $r\in[2M,\infty]$.
\item
The radial tension $\tau$ is negative in the range $r\in
(2.3M,\infty]$, and positive in the range
$r\in[2M,2.3M)$.
\item
The transverse pressure $p$ is negative over the entire
range $r\in[2M,\infty]$.
\end{itemize}
While the analytic approximation and numeric data disagree on where
the bumps and zero-crossings are located, there is good overall
agreement as to the qualitative shape of these curves. In this
paper I will need to use only relatively crude aspects of the
numeric data and eyeball inspection is quite sufficient. For instance
from~\cite[figure 4, page 3005]{JLO92} it is clear that
\begin{itemize}
\item 
$-\tau < \rho$ outside the horizon. Hence $\rho+\tau<0$.
\item
$p< -\tau$ outside the horizon. Hence $\tau+p<0$.
\item
$\rho<p$ outside the horizon. Hence $\rho-p<0$.
\end{itemize}
Slightly more subtle are the following relationships, also derivable
by visual inspection
\begin{itemize}
\item 
$|\tau| < -\rho$ outside the horizon. Hence $\rho+|\tau|<0$. Whence
$\rho\pm\tau<0$.
\item
$|\tau|<-p$ outside the horizon. Hence $p+|\tau|<0$.  Whence
$p\pm\tau<0$.
\end{itemize}
This will be sufficient for current purposes.

%%%%%%%%%%%%%%%%%%%%%%%%%%%%%%%%%%%%%%%%%%%%%%%%%%%%%%%%%%%%%%%%%%%%%%%%%%%%%% 
\section{Point-wise energy conditions} 
%%%%%%%%%%%%%%%%%%%%%%%%%%%%%%%%%%%%%%%%%%%%%%%%%%%%%%%%%%%%%%%%%%%%%%%%%%%%%% 
\subsection{Outside the horizon:} 
%%%%%%%%%%%%%%%%%%%%%%%%%%%%%%%%%%%%%%%%%%%%%%%%%%%%%%%%%%%%%%%%%%%%%%%%%%%%%% 
 
Outside the event horizon, the NEC reduces to the pair of constraints 
\begin{equation} 
\rho(r) - \tau(r) \geq 0? \qquad \rho(r) + p(r) \geq 0? 
\end{equation} 
But we have already seen that both $\rho$ and $p$ are individually
negative everywhere outside the event horizon, the numeric data
and the analytic approximation agreeing on this point.

Therefore the NEC is definitely violated everywhere outside the
event horizon.  This automatically implies that all the other
point-wise energy conditions (WEC, SEC, and DEC) are violated
outside the event horizon.

For future use I point out that (defining $z=2M/r$ to reduce
clutter) the analytic approximation gives
\begin{eqnarray} 
\rho(z)-\tau(z) &=&  
-4 \; p_\infty \; z^6 \; 
{[28 - 48 z + 21z^2]  \over (1 - z)^2 },
\\
\rho(z) + p(z) &=& 
- 4 \; p_\infty \; z^6 \;
{[34 - 60z +27z^2] \over (1 - z)^2 }.
\end{eqnarray}
It is easy to verify that both of these expressions are strictly
negative outside the event horizon, and indeed are strictly negative
throughout the spacetime. 

This completes the proof of Theorem 1.

%%%%%%%%%%%%%%%%%%%%%%%%%%%%%%%%%%%%%%%%%%%%%%%%%%%%%%%%%%%%%%%%%%%%%%%%%%%%%% 
\subsection{Inside the horizon:} 
%%%%%%%%%%%%%%%%%%%%%%%%%%%%%%%%%%%%%%%%%%%%%%%%%%%%%%%%%%%%%%%%%%%%%%%%%%%%%% 
 
Inside the event horizon, the radial coordinate becomes timelike, 
and the roles played by $\rho(r)$ and $\tau(r)$ are interchanged. 
The NEC reduces to the pair of constraints 
\begin{equation} 
\tau(r) -\rho(r)\geq 0? \qquad \tau(r) + p(r) \geq 0? 
\end{equation} 
 
We now only have the analytic approximation available. We have
already seen above that $\tau-\rho>0$, so this condition is not
going to help us. On the other hand
\begin{eqnarray} 
\tau(z)+p(z) &=&  - 24 \; \; p_\infty z^6,
\end{eqnarray}
which is blatantly negative inside the event horizon (and indeed
throughout the spacetime).

Therefore, {\em assuming the analytic approximation is not wildly
inaccurate}, the NEC is violated everywhere inside the
event horizon, and consequently all the other point-wise energy
conditions (WEC, SEC, and DEC) are violated as well. 

In summary, all the point-wise energy conditions are  violated
throughout the entire Schwarzschild spacetime. Outside the event
horizon we have both numeric data and analytic approximations which
agree on this point. Inside the event horizon we have only the
analytic approximation. 

So far, what I have shown is at that at each point in the spacetime
there is at least one null/timelike vector along which the pointwise
energy conditions are violated.  But it is possible to derive  much
stronger results.

%%%%%%%%%%%%%%%%%%%%%%%%%%%%%%%%%%%%%%%%%%%%%%%%%%%%%%%%%%%%%%%%%%%%%%%%%%%%%% 
\section{Total external NEC violation} 
%%%%%%%%%%%%%%%%%%%%%%%%%%%%%%%%%%%%%%%%%%%%%%%%%%%%%%%%%%%%%%%%%%%%%%%%%%%%%% 
 
%%%%%%%%%%%%%%%%%%%%%%%%%%%%%%%%%%%%%%%%%%%%%%%%%%%%%%%%%%%%%%%%%%%%%%%%%%%%%% 
\subsection{Outside the horizon:} 
%%%%%%%%%%%%%%%%%%%%%%%%%%%%%%%%%%%%%%%%%%%%%%%%%%%%%%%%%%%%%%%%%%%%%%%%%%%%%% 
 
Consider a generic null vector inclined at an angle $\psi$ away
from the radial direction.  Then without loss of generality, in an
orthonormal frame attached to the $(t,r,\theta,\phi)$ coordinate
system,
\begin{equation} 
k^{\hat \mu} \propto (\pm1,\cos\psi, 0, \sin\psi). 
\end{equation} 
Ignoring the (presently irrelevant) overall normalization of 
the null vector, one is interested in calculating 
\begin{eqnarray} 
\langle B | T_{\mu\nu} | B \rangle \; k^\mu k^\nu  
&\propto& (\rho -\tau \cos^2\psi + p \sin^2\psi)  
\nonumber\\ 
&=& 
([\rho -\tau]  + [\tau +p ]\sin^2\psi). 
\end{eqnarray} 
I intend to show that this quantity is negative for all values of
$\psi$ and $r$.

We have already seen that the analytic approximation implies that
$\rho-\tau$ is negative outside the event horizon (and in fact is
negative throughout the spacetime).  We have also seen that this
observation can be extended to the numerical data by inspection of
the graphs plotted in~\cite{JLO92}.

For the analytic approximation we have also seen that $\tau(z)+p(z)$
is blatantly negative outside the event horizon (and indeed throughout
the spacetime). Furthermore, we have already seen that this observation
can be extended to the numerical data by inspection of the graphs
plotted in~\cite{JLO92}.

We are now done. (Both terms in square brackets are strictly
negative). What we have shown is that for any point $p$ outside
the event horizon, and any null vector $k$ the inner product $\langle
T_{\mu\nu}\rangle \; k^\mu k^\nu$ is strictly negative.  This
completes the proof of Theorem 2.

%%%%%%%%%%%%%%%%%%%%%%%%%%%%%%%%%%%%%%%%%%%%%%%%%%%%%%%%%%%%%%%%%%%%%%%%%%%%%% 
\subsection{Inside the horizon:} 
%%%%%%%%%%%%%%%%%%%%%%%%%%%%%%%%%%%%%%%%%%%%%%%%%%%%%%%%%%%%%%%%%%%%%%%%%%%%%% 
 
Inside the event horizon there are additional technical
fiddles. One should now consider a generic null vector inclined at
an angle $\tilde\psi$ away from the $t$ direction (which is now
spacelike). Then without loss of generality, in an orthonormal
frame attached to the $(t,r,\theta,\phi)$ coordinate system,
\begin{equation} 
k^{\hat \mu} \propto (\cos\tilde\psi,\pm 1,0,\sin\tilde\psi). 
\end{equation} 
One should now consider the quantity 
\begin{eqnarray} 
\langle B | T_{\mu\nu} | B \rangle \; k^\mu k^\nu  
&\propto& 
(\tau -\rho\cos^2\tilde\psi + p\sin^2\tilde\psi) 
\nonumber\\ 
&\propto& 
(-[\rho-\tau]  + [\rho +p ]\sin^2\tilde\psi). 
\end{eqnarray} 

We are now limited to the analytic approximation, and (subject to
suitable caveats) have already seen that inside the horizon
$\rho-\tau$ and $\rho+p$ are both everywhere negative.

Because of the relative minus sign the critical issue is now the
relative magnitude of the terms $|\rho-\tau|$ and $|\rho+p|$. The
quantity $\langle T_{\mu\nu}\rangle \; k^\mu k^\nu$ can be made
positive by choosing:
\begin{equation}
\sin^2\tilde\psi < \sin^2(\tilde\psi_{\rm crit}(z)) \equiv
{ [28 - 48 z + 21z^2]\over [34-60z+27z^2]}.
\end{equation}
 
Just inside the horizon,  $z=1^-$, one has $\tilde\psi_{\rm
crit}(z=1^-) = (\pi/2)^-$, with all the stress-energy components
diverging as one actually hits the horizon.  As one approaches the
singularity for $z$ large one has $\tilde\psi_{\rm crit}(z) \to
\sin^{-1}(7/9)\approx 62 \deg$.

In summary: Inside the event horizon the analytic approximation
suggests that certain null directions allow one to have $\langle
T_{\mu\nu}\rangle \; k^\mu k^\nu>0$. This is the situation for
which I introduced the notion of the partial null energy condition
(PNEC) in reference~\cite{Visser96a}.

In some sense this is the mirror image of the situation in the
Hartle--Hawking vacuum. In that vacuum I found that the NEC was
satisfied at large radius, and that at small radius it was possible
to find certain directions such that PNEC was violated. Here in
the Boulware vacuum I find that NEC is violated at large radius,
and that at small radius it is possible to find certain directions
such that PNEC  is satisfied.

%%%%%%%%%%%%%%%%%%%%%%%%%%%%%%%%%%%%%%%%%%%%%%%%%%%%%%%%%%%%%%%%%%%%%%%%%%%%%% 
\subsection{Total external WEC/DEC violation.} 
%%%%%%%%%%%%%%%%%%%%%%%%%%%%%%%%%%%%%%%%%%%%%%%%%%%%%%%%%%%%%%%%%%%%%%%%%%%%%% 

It is now a simple exercise to extend this type of analysis to
generic timelike vectors. Outside the horizon one can take
\begin{equation} 
V^{\hat \mu} = \gamma(\pm1,\beta \cos\psi, 0, \beta\sin\psi). 
\end{equation} 
The quantity of interest is now
\begin{eqnarray} 
\langle B | T_{\mu\nu} | B \rangle \; V^\mu V^\nu  
&=& \gamma^2( \rho -\beta^2\tau \cos^2\psi + \beta^2 p \sin^2\psi)  
\nonumber\\ 
&=& 
\gamma^2([\rho -\beta^2\tau]  + \beta^2[\tau +p ]\sin^2\psi). 
\end{eqnarray} 
Both of the quantities in square brackets are everywhere negative
outside the event horizon. Note that $\rho -\beta^2\tau < \rho
+\beta^2|\tau| < \rho + |\tau| < {\rm Max}(\rho +\tau, \rho-\tau)
<0$. Consequently\\
%%%\begin{theorem}[Total external WEC violation]\hfil\break
{\bf Theorem 3 [Total external WEC violation]}\hfil\break
{\em
For a conformally-coupled massless scalar field on a Schwarzschild
spacetime in the Boulware vacuum state:  For {\em any} point $p$
outside the event horizon and {\em any} timelike vector $V$
\begin{equation}
\langle B | T^{\mu\nu} | B \rangle \; V_\mu V_\nu \; \leq 0.
\end{equation}
The equality is in fact achieved only at spatial infinity.
}
%%%\end{theorem}

Observe that while the violations of the ordinary NEC immediately
imply violations of the ordinary WEC, there is something extra to
be proved here when one wants to discuss the wholesale, everywhere
in the phase space, violations addressed in this paper.

The total WEC violation theorem now immediately implies\\
%%%\begin{theorem}[Total external DEC violation]\hfil\break
{\bf Theorem 4 [Total external DEC violation]}\hfil\break
{\em
For a conformally-coupled massless scalar field on a Schwarzschild
spacetime in the Boulware vacuum state:  For {\em any} point $p$
outside the event horizon and {\em any} timelike vector $V$ the
dominant energy condition is violated.
}
%%%\end{theorem}

Turning attention to the region inside the event horizon one can
take
\begin{eqnarray} 
V^{\hat \mu} = \gamma(\beta \cos\tilde\psi,\pm1, 0, \beta\sin\tilde\psi). 
\end{eqnarray} 
The quantity of interest is
\begin{eqnarray} 
\langle B | T_{\mu\nu} | B \rangle \; V^\mu V^\nu  
&=& 
\gamma^2(\tau -\beta^2\rho\cos^2\tilde\psi + \beta^2 p\sin^2\tilde\psi) 
\nonumber\\ 
&=& 
\gamma^2([\tau-\beta^2\rho]  + \beta^2[\rho+p ]\sin^2\tilde\psi). 
\end{eqnarray} 
While $\rho+p$ is everywhere negative it is relatively easy to
drive the total positive: For instance take $\beta=0$ and
$r\in(1.7753M,2M)$.  (We have already seen that $\tau$ is positive
in this range.) Alternatively one can take $\beta\approx1$ and
recover the NEC discussion.

In summary: Inside the event horizon there are certainly some points
and some timelike vectors for which the analytic approximation
suggests $\langle T_{\mu\nu} \rangle \; V^\mu V^\nu >0$.

%%%%%%%%%%%%%%%%%%%%%%%%%%%%%%%%%%%%%%%%%%%%%%%%%%%%%%%%%%%%%%%%%%%%%%%%%%%%%% 
\subsection{Total external SEC violation.} 
%%%%%%%%%%%%%%%%%%%%%%%%%%%%%%%%%%%%%%%%%%%%%%%%%%%%%%%%%%%%%%%%%%%%%%%%%%%%%% 

Finally we turn to the issue of wholesale violations of the SEC.
If SEC were to hold one would wish to prove
\begin{equation}
\langle B | {\bar T}^{\mu\nu} | B \rangle \; V_\mu V_\nu \; \geq 0?
\end{equation}
Here $\bar T$ is the trace-reversed stress tensor 
\begin{equation}
{\bar T}^{\mu\nu}
= T^{\mu\nu} - {1\over2} g^{\mu\nu} T. 
\end{equation}
(The easiest way to remember exactly what the SEC is means is to
think of it as the trace-reversed WEC.) Thus outside the horizon
we can repeat the analysis used for the WEC by making the substitutions
\begin{eqnarray}
  \rho&\to&\bar\rho      = (\rho-\tau+2p)/2, \nonumber\\
  \tau&\to&\bar\tau      = (\tau-\rho+2p)/2, \nonumber\\
     p&\to&\bar p        = (\rho+\tau)/2.
\end{eqnarray}
For instance, outside the horizon the quantity of interest becomes
%\widetext
\begin{eqnarray} 
\langle B | {\bar T}_{\mu\nu} | B \rangle \; V^\mu V^\nu  
&=& 
\gamma^2([\bar\rho -\beta^2\bar\tau]  + \beta^2[\bar\tau +\bar p ]\sin^2\psi)
\nonumber\\
&=&
\gamma^2\{(1+\beta^2)[\rho -\tau]/2 + (1-\beta^2)[p]  
\nonumber\\
&&\qquad 
+ \beta^2[\tau + p ]\sin^2\psi\}.  
\end{eqnarray} 
%\narrowtext
You will by now be unsurprised at the refrain: each quantity in
square brackets is everywhere negative outside the event horizon,
the analytic approximation and the numerical data agreeing on this
point. That is:\\
%%%\begin{theorem}[Total external SEC violation]\hfil\break
{\bf Theorem  5 [Total external SEC violation]}\hfil\break
{\em
For a conformally-coupled massless scalar field on a Schwarzschild
spacetime in the Boulware vacuum state:  For {\em any} point $p$
outside the event horizon and {\em any} timelike vector $V$
\begin{equation}
\langle B | {\bar T}^{\mu\nu} | B \rangle \; V_\mu V_\nu \; \leq 0.
\end{equation}
The equality is in fact achieved only at spatial infinity.
}
%%%\end{theorem}

If we now look at the region inside the horizon the relevant quantity
is
%\widetext
\begin{eqnarray} 
\langle B | {\bar T}_{\mu\nu} | B \rangle \; V^\mu V^\nu  
&=& 
\gamma^2 
(\bar\tau - 
\beta^2 \bar\rho \cos^2\tilde\psi  + 
\beta^2 \bar p   \sin^2\tilde\psi)
\nonumber\\
&=& 
\gamma^2 
([\bar\tau-\beta^2\bar\rho]  + \beta^2[\bar\rho + \bar p ]\sin^2\tilde\psi)
\nonumber\\
&=&
\gamma^2 
\{-(1+\beta^2)[\rho-\tau]/2-(1-\beta^2)[p]
\nonumber\\
&&\qquad 
+ \beta^2[\rho+p]\sin^2\tilde\psi\}
\end{eqnarray} 
%\narrowtext
In the last line $\rho-\tau$, $p$, and $\rho+p$ are individually negative, 
but the relative minus signs make it easy to drive this quantity positive.
Note that if one takes $\beta\to1$, one recovers the NEC discussion.

In summary: Inside the event horizon there are certainly some points
and some timelike vectors for which the analytic approximation
suggests $\langle \bar T_{\mu\nu} \rangle \; V^\mu V^\nu >0$.

%%%%%%%%%%%%%%%%%%%%%%%%%%%%%%%%%%%%%%%%%%%%%%%%%%%%%%%%%%%%%%%%%%%%%%%%%%%%%% 
\section{Exotic energy conditions} 
%%%%%%%%%%%%%%%%%%%%%%%%%%%%%%%%%%%%%%%%%%%%%%%%%%%%%%%%%%%%%%%%%%%%%%%%%%%%%% 

The total NEC violation theorem proved above is, at least in the
region outside the event horizon,  strong enough to kill all
conditional energy conditions such as the PNEC and Scri--NEC
introduced in~\cite{Visser96a}.

Furthermore, it is strong enough to kill any and all energy conditions
constructed by averaging the quantity $\langle T_{\mu\nu} \rangle
\; k^\mu k^\nu$ any sort of null curve and demanding positivity
for the resulting integral:
\begin{itemize}
\item 
The ANEC is violated along all null geodesics that do not cross
the event horizon, and in fact is also violated along all such
non-geodesic null curves.
\item 
Smearing the ANEC by averaging in transverse
directions~\cite{Flanagan-Wald} will not help. The smeared ANEC
will still be violated along any and all null curves that avoid
the event horizon.
\item
Semi-local versions of ANEC, obtained by inserting any arbitrary
weighting function $f(\lambda)$ into the ANEC integral, and demanding
that the integral remain positive, are also killed by this result.
\end{itemize}

Similarly, ``total WEC violation theorem'' and ``total SEC violation
theorem''  guarantee that the averaged weak energy condition (AWEC)
and averaged strong energy condition (ASEC) and their variants are
also guaranteed to be violated for all timelike curves, geodesic
or not, with arbitrary weighting functions, provided only they
avoid the region behind the event horizon.

On the other hand, the ``quantum inequalities'' of Ford and
Roman~\cite{Ford-Roman93,Ford-Roman96,Ford-Roman95b} are {\em not}
necessarily violated by these results. Extending the analysis of
Ford and Roman, it seems that for timelike curves in a non-flat
spacetime the generalized quantum inequalities would take the
generic form
\begin{equation}
\int_\gamma f(\tau) \; \langle T^{\mu\nu} \rangle \; V_\mu V_\nu \; d\tau
\; \geq \; - | Q[f,g] |.
\end{equation}
Here $f(\tau)$ is some specific weighting function, and $Q[f,g]$ is
some functional of the weighting function and the spacetime metric.
The ``quantum inequality'' states that this $f$-weighted AWEC is
not allowed to become excessively negative.  (But because it {\em
is} allowed to become negative the quantum inequalities are compatible
with the results of this paper.)

It would clearly be of interest to consider more general weighting
functions than the specific choice made by Ford and Roman, and
would also be very interesting to see to what extent one can obtain
singularity theorems or positive mass theorems based on such
generalized quantum inequalities.

%%%%%%%%%%%%%%%%%%%%%%%%%%%%%%%%%%%%%%%%%%%%%%%%%%%%%%%%%%%%%%%%%%%%%%%%%%%%%% 
\section{Discussion} 
%%%%%%%%%%%%%%%%%%%%%%%%%%%%%%%%%%%%%%%%%%%%%%%%%%%%%%%%%%%%%%%%%%%%%%%%%%%%%% 
 
In a companion paper~\cite{Visser96a} I have studied the Hartle--Hawking
vacuum state, discovering a complicated layering of energy-condition
violations confined to the region between the unstable photon orbit
and the event horizon.

The situation in the Boulware vacuum is more dramatic: 
\begin{itemize}
\item
{\bf (1) All standard (point-wise and averaged) energy conditions
are violated throughout the entire region exterior to the event
horizon. }
\item
{\bf (2) Outside the event horizon, the standard point-wise energy
conditions are violated in a maximal manner: they are violated at
all points and for all null/timelike vectors.}
\item
{\bf (3) The standard point-wise energy conditions seem to be
violated even inside the event horizon.}
\end{itemize}

It should be borne in mind that these are test-field limit
calculations, which gives us a (mild) excuse to not worry too much.
Furthermore, the Boulware vacuum is ill-behaved on the event horizon
itself, and for this reason it might be thought to be an ``unphysical''
quantum state, giving a further excuse for not worrying.

But this is not the whole story: the Boulware vacuum is believed
to be a good approximation to the quantum mechanical vacuum
surrounding a large condensed object such as a star or planet that
has not been allowed to collapse past its Schwarzschild radius.
Because the mode-sums and subtractions used in calculating $\langle
T \rangle$ are purely local both the analytic approximations and the
numerical calculations should be perfectly adequate for describing
the vacuum polarization outside the central body itself.

(There is a potential subtlety here: Outside the central object
the modes are simply given by the solutions to the Regge--Wheeler
equation, and so are determined in a purely local manner. On the
other hand, properly determining the overall normalization of each
mode depends on an integral over an entire Cauchy surface. This is
where non-local effects might sneak in. It is thus conceivable,
though maybe unlikely, that the vacuum polarization outside a star
or planet could depend on details of its interior composition. On
the other hand, one still expects the analytic approximation
discussed in this paper to be a rather good approximation outside
the central body---and the analytic approximation is blatantly
local.)

The analysis of this note suggests that the entire region outside
the central body should violate all the standard energy conditions.
These violations will be tiny to be sure, but they will be there
in the test-field limit. It is of considerable interest to provide
even a single quantum state that leads to such wholesale violations
of the energy conditions.

Now it is conceivable that this effect would go away if one were able
to find a fully self-consistent solution to the field equations of
semiclassical quantum gravity---this is a very interesting question
well beyond the scope of this paper.

However some initial steps towards self-consistency can be made by
taking a perturbative point of view. Consider perturbative
self-consistency in the sense of Flanagan and Wald~\cite{Flanagan-Wald},
and view the spacetime as being described by a class of metrics
$g_{\mu\nu}(x,\epsilon)$, where $\epsilon$ is to be thought of as
a perturbation parameter. [The relevant expansion parameter is in
fact $\epsilon = \hbar/M^2 = (m_P/M)^2$, and is simply the square
of the ratio of the mass of the central body to the Planck mass.]

We want to take $\epsilon=0$ to correspond to the Schwarzschild
metric, calculate the gravitational polarization in the Boulware
metric (which by definition is of order $\epsilon$), and feed this
back in to get the first-order shifted metric $g_{\mu\nu}(x,\epsilon)=
g_{\mu\nu}(x,0) + \epsilon \Delta g_{\mu\nu}(x) + O(\epsilon^2)$.
Then this first-order shifted metric has a vacuum polarization
which is equal to that of the Schwarzschild geometry---up to first
order in $\epsilon$---and thus provides a first-order self-consistent
solution of semiclassical quantum gravity.  Given the smallness of
$\epsilon$ for heavy objects we should expect the first-order
self-consistent solution to be extremely close to the exact solution.

The fly in the ointment is that this self-consistent solution is
breaking down at $r=2M$, the location of the event horizon in zero-th
order, which is a much more confusing place at first order. Be that
as it may, one may think of a star or planet and chop the
almost-Schwarzschild geometry off at some suitably large radius,
matching it to some stellar or planetary interior.

Outside the star or planet the entire analysis of this paper should
hold, at least qualitatively,  and we would then have a self-consistent
solution of semiclassical quantum gravity that violates all of the
energy conditions. In the notation of Flanagan and
Wald~\cite{Flanagan-Wald}, $I^{(1)}$, the first-order perturbation
of the ANEC integral will be negative for geodesics that remain in
this region. Because I was able to prove that ANEC was violated at
this order for all null curves it is clear that the transverse
averaging advocated by Flanagan and Wald will not help the situation.

Thus I claim that the results of this note are highly suggestive
evidence pointing to the fact that general self-consistent solutions
of semiclassical quantum gravity will {\em not} satisfy the
energy conditions, and may in fact for certain quantum fields and
certain quantum states violate {\em all} the standard energy conditions.

%%%%%%%%%%%%%%%%%%%%%%%%%%%%%%%%%%%%%%%%%%%%%%%%%%%%%%%%%%%%%%%%%%%%%%%%%%%%%%
\acknowledgements
%%%%%%%%%%%%%%%%%%%%%%%%%%%%%%%%%%%%%%%%%%%%%%%%%%%%%%%%%%%%%%%%%%%%%%%%%%%%%%

I wish to thank Nils Andersson, \'Eanna Flanagan, Larry Ford, and
Tom Roman for their comments and advice.

The numerical analysis in this paper was carried out with the aid
of the Mathematica symbolic manipulation package.
    
This research was supported by the U.S. Department of Energy. 
 
%%%%%%%%%%%%%%%%%%%%%%%%%%%%%%%%%%%%%%%%%%%%%%%%%%%%%%%%%%%%%%%%%%%%%%%%%%%%%% 
%%%%%%%%%%%%%%%%%%%%%%%%%%%%%%%%%%%%%%%%%%%%%%%%%%%%%%%%%%%%%%%%%%%%%%%%%%%%%% 
 
%%%%%%%%%%%%%%%%%%%%%%%%%%%%%%%%%%%%%%%%%%%%%%%%%%%%%%%%%%%%%%%%%%%%%%%%%%%%%% 
\end{document}